\documentclass[prd,aps,twocolumn,superscriptaddress,preprintnumbers,nofootinbib,showpacs]{revtex4}
\usepackage{graphicx}
\usepackage{exscale}
\usepackage[intlimits]{amsmath}
\usepackage{amsfonts}
\usepackage{amssymb,amscd}
\usepackage{epsfig}
\usepackage{pstricks}

\newcounter{commentdepth}
\newcommand{\be}{\begin{equation}}
\newcommand{\ee}{\end{equation}}

\newcommand{\beqa}{\begin{eqnarray}}
\newcommand{\eeqa}{\end{eqnarray}}
\newcommand{\beq}{\begin{equation}}
\newcommand{\eeq}{\end{equation}}

\newcommand{\p}{\partial}

\newcommand{\reftitle}[1]{}

\begin{document}
%%\preprint{\parbox[t]{\textwidth}
%%{\small May 26, 2008 \hspace*{5mm} \#\#\# \hfill 
%%ArXiv:0805.3804 [hep-ph]}}

\title{A model of color confinement\footnote{Talk given at Quark Confinement and Hadron Spectrum IX, Madrid, Spain, Aug. 30 to Sept. 3, 2010.}}

%\author{Klaus~Lichtenegger}%
%\affiliation{Karl-Franzens-Universit\"at Graz, 8010 Graz, Austria}%

\author{Daniel~Zwanziger}
\affiliation{Physics Department, New York University, New York, NY 10003, USA}

\begin{abstract}
\noindent   	

	A simple model is presented that describes the free energy $W(J)$ of QCD coupled to an external current that is a single plane wave, $J(x) = H \cos(k \cdot x)$.  The model satisfies a bound obtained previously on $W(J)$ that comes from the Gribov horizon.  If one uses this model to fit recent lattice data --- which give for the gluon propagator $D(k)$ a non-zero value, $D(0) \neq 0$, at $k = 0$ --- the data favor a non-analyticity in $W(J)$. 

\end{abstract}


\maketitle

\section{Introduction}

	Recent numerical studies on large lattices of the gluon propagator $D(k)$ in Landau gauge in 3 and 4 Euclidean dimensions, reviewed recently in \cite{Cucchieri:2010.03}, yield finite values for $D(0) \neq 0$ \cite{Cucchieri:2008.12} - \cite{Cucchieri:2007b}, in apparent disagreement with the theoretical expectation that $D(0) = 0$, originally obtained by Gribov~\cite{Gribov:1977wm}, and argued in~\cite{Zwanziger:1991}.  Upon reviewing the argument~\cite{Zwanziger:1991} which leads to $D(0) = 0$, one hypothesis stands out which should perhaps be dropped in view of the apparent disagreement.  This is the hypothesis that the free energy $W(J)$ in the presence of sources $J$ is analytic in $J$.  This is an important point because a non-analyticity in the free energy is characteristic of a change of phase.  

	The free energy $W(J)$ enters the picture because it is the generating function of the connected gluon correlators.  In particular the gluon propagator $D_{x,y}$, is a second derivative of $W(J)$ at $J = 0$,
\beq
D_{x, y} = {\p^2 W(J) \over \p J_x \p J_y}\Big|_{J=0}.
\eeq
Here a condensed index notation is used, where $J_x$ represents $J_\mu^b(x)$, $\mu$ is a Lorentz index, and $b$ is a color index, and we write
\beq
(J, A) = \sum_x J_x A_x =  \int d^Dx \ J_\mu^b(x) A_\mu^b(x).
\eeq
The free energy $W(J)$ in the presence of sources $J$ is given by
\beqa
\exp W(J) & = & \langle \exp(J, A) \rangle
\nonumber \\
& = & \int_\Omega dA \ \rho(A) \exp(J, A),
\eeqa
where $\rho(A)$ is a positive, normalized probability distribution.  The integral over $A$ is effected in Landau gauge $\p_\mu A_\mu = 0$.  The domain of integration is restricted to the Gribov region $\Omega$, a region in $A$-space where the Faddeev-Popov operator is non-negative, $M(A) \equiv - \p_\mu D_\mu(A) \geq 0$.

	The model will be defined for the special case where the source is given by
\beq
J_\mu^b(x) = H_\mu^b \cos(k \cdot x),
\eeq
so the free energy
\beq
\exp W_k(H) =
\langle \ \exp[ \sum_x H_\mu^b \cos(k \cdot x) A_\mu^b(x) ] \ \rangle,
\eeq
depends only on the parameters $k_\mu$ and $H_\mu^b$.  Because $A_\mu(x)$ is transverse, only the transverse part oif $H$ is operative, and we impose
\beq
k_\mu H_\mu^b = 0.
\eeq  
By analogy with spin models, $H_\mu^b$ may be interpreted as the strength of an external magnetic field, with a color index $b$, which is modulated by a plane wave $\cos(k \cdot x)$.   

	A rigorous bound for $W_k(H)$ on a finite lattice is given in \cite{Zwanziger:1991}.  One can easily show that  in the limit of large lattice volume $V$, and in the continuum limit, this implies the Lorentz-invariant continuum bound in $D$ Euclidean dimensions,
\beq
\label{LIcontbound}
w_{k}(H) \equiv {W_k(H) \over V} \leq (2 D k^2)^{1/2} |H|,
\eeq
where $|H|^2 = \sum_{\mu, b} (H_\mu^b)^2$ is the color- and Lorentz-invariant norm of $H_\mu^b$.  [In our notation the vector potential is given by $A(x) = g A^{\rm pert}(x)$, and has engineering dimension in mass units $[A(x)] = 1$ in all Euclidean dimension $D$, while $[H] = D-1$.]

	This bound yields in the zero-momentum limit
\beq
w_{0}(H) = \lim_{k \to 0} w_{k}(H) = 0.
\eeq
As discussed in \cite{Zwanziger:1991}, this states that the system does not respond to a constant external color-magnetic field {\em no matter how strong}.  In this precise sense, the color degree of freedom is absent, and color is confined.

\section{The model}

	The model of color confinement is defined by the expression for the free energy
\beq
\label{model}
W_k(H) = c |k| V \{ [H^2 + f^2(k)]^{1/2} - f(k) \},
\eeq	
where $c$ is a dimensionless constant, and $f(k) \geq 0$ is an as yet undetermined function.  For $c = (2D)^{1/2}$ this expression for $W_k(H)$ satisfies the bound (\ref{LIcontbound}), as one sees from $(H^2 + f^2)^{1/2} \leq |H| + f$.  The bound (\ref{LIcontbound}) is saturated for small $k$ if
\beq
f(0) = 0.
\eeq
For example we shall later take
\beq
\label{fofk}
f(k) = m^{D-2} |k|,
\eeq
where $m$ is a mass.  

	One can easily verify that expression (\ref{model}) also satisfies $W_k(0) = 0$, as required for normalized probability $\rho(A)$, and that the gluon propagator, defined below, is a positive matrix, as required. 
	
	By definition the ``classical field" is the expectation-value 
\beq
a_{\mu}^b(k ,H) \equiv \langle a_\mu^b(k) \rangle|_H
\eeq
of the field
\beq
a_{\mu}^b(k) \equiv \int_V d^Dx \ \cos(k \cdot x) A_\mu^b(x)
\eeq
in the presence of the source $H$, and at momentum $k$.  It is expressed in terms of $W_k(H)$, by
\beq
\label{asubkofH}
a_\mu^b(k, H) \equiv {\p W_k(H) \over \p H_\mu^b} = c |k| V { H_\mu^b \over [H^2 + f^2(k)]^{1/2}},
\eeq
and the gluon propagator by
\beqa
\label{propagator}
D_{\mu \nu}^{bc}(k, H) & \equiv & V^{-1} {\p^2 W_k(H) \over \p H_\mu^b \p H_\nu^c}
\\    \nonumber
 & = & c |k| {P_{\mu \nu}^T(k) \delta^{bc}[H^2 + f^2(k)] + H_\mu^b H_\nu^c \over [H^2 + f^2(k)]^{3/2}}.
\eeqa
Here $P_{\mu \nu}^T(k) \equiv \delta_{\mu \nu} - k_\mu k_\nu / k^2$ is the transverse projector which appears because $H_\mu^b$ is identically transverse $k_\mu H_\mu^b = 0$, and $\p H_\mu^b / \p H_\nu^c = P_{\mu \nu}^T (k)\delta^{bc}$.

	The free energy $W_k(H)$ vanishes at $k = 0$ for all $H$,
\beq
W_0(H) = 0.
\eeq	
If $W_k(H)$ were analytic in $H$ in the limit $k \to 0$, this would imply that all derivatives of $W_0(H)$ vanish and with it the gluon propagator (\ref{propagator}) at $k = 0$, $D(0) = 0$.  However this disagrees with recent lattice data~\cite{Cucchieri:2010.03} which give a finite result, $D(0) \neq 0$, in Euclidean dimensions 3 and 4.  However the second (and higher) derivative $W_k''(H)$ is non-analytic in $H$ in the limit $k \to 0$ when $f(0) = 0$, and the gluon propagator (\ref{propagator}) 
\beq
D_{\mu \nu}^{bc}(k) = D_{\mu \nu}^{bc}(k, 0) = { c |k| \over f(k) } P_{\mu \nu}^T \delta^{bc}
\eeq
does not necessarily vanish for $k \to 0$.  Indeed with $f(k) = m^{D-2} |k|$,
we have
\beq
D(0) = c/m^{D-2},
\eeq 
which is finite.  Thus our model, with $f(k) = m^{D-2} |k|$ satisfies the bound (\ref{LIcontbound}) and accords with lattice data for $D = 3$ and $D = 4$.   In contradistinction to the previous treatment \cite{Zwanziger:1991}, the hypothesis that $W_k(H)$ is analytic in $H$ for $k \to 0$ is relaxed.  Moreover, for $f(k) = m^{D-2+ \alpha} |k|^{1-\alpha}$, we get 
\beq
D(k) = {c k^{\alpha} \over m^{D-2 + \alpha} },
\eeq  
which fits the data in dimension $D = 2$ for $\alpha = 1/5$ \cite{Maas:2007}.

	If one fits the lattice data with this model, the data favor the value $f(0) = 0$ in $D = 2, 3,$ and $4$ dimensions.  With the value $f(0) = 0$, the model gives asymptotically at low $k$
\beq
W_k(H) \approx W_k^{\rm as}(H) = c |k| |H| V.
\eeq
This is linear in $|H|$, whereas normally one expects the free energy $W_k( H)$ to be a power series in $H$, with leading term of order $H^2$.

\section{Model quantum effective action}

	As in statistical mechanics we define, for each momentum $k$, the analog of the bulk magnetization
\beq
M_\mu^b(k, H) = {\p W_k(H) \over \p H_\mu^b},
\eeq		
and make the Legendre transformation from $W_k(H)$ to
\beq
\label{legendreM}
\Gamma_k(M_k) = M_kH - W_k(H).
\eeq	
In fact the ``magnetization" $M_k(H)$, coincides in our gauge theory with the classical gluon field given in (\ref{asubkofH}),
\beq
a_k(H) = M_k(H),
\eeq 	
and the Legendre transformation (\ref{legendreM}) yields the quantum effective action of the gauge theory
\beq
\Gamma_k(a_k) = a_kH - W_k(H),
\eeq
More precisely this is the quantum effective action with all variables set to 0 except $a_\mu^b(k)$ for fixed $k$.  To find $\Gamma_k(a_k)$ we note that
\beqa
\Gamma_k(H) & = & { c|k|V H^2 \over [H^2 + f^2(k)]^{1/2} } - W_k(H) 
\nonumber    \\
& = & c|k| f(k) V \Big\{ 1 - {f(k) \over [H^2 + f^2(k)]^{1/2} } \Big\},
\eeqa
and from squaring (\ref{asubkofH}) we find
\beq
{ f \over [H^2 + f^2(k)]^{1/2} } = \Big[1 - \Big({a_k \over c|k|V } \Big)^2\Big]^{1/2},
\eeq
which gives for the quantum effective action with all variables except $a_k$ set to 0
\beq
\label{QEF}
\Gamma_k(a_k) = c |k| f(k) V \Big\{ 1 - \Big[1 - \Big({|a_k| \over c|k|V } \Big)^2\Big]^{1/2} \Big\}.
\eeq
It is singular at
\beq
|a(k)| = c |k| V.
\eeq

\bigskip
	  
{\bf Acknowledgements}\\
The author recalls with pleasure stimulating conversations with Attilio Cucchieri and Axel Maas.

%%%%%%%%%%%%%%%%%%%%%%%%%%%%%%%%%%%%%%%%

%%%%%%%%%%%%%%%%%%%%%%%%%%%%%%%%%%%%%%%%%%%%%%%%%%%	


\begin{thebibliography}{99}
\parskip=0pt

% papers cited in the main article
%%%%%%%%%%%%%%%%%



\bibitem{Cucchieri:2010.03}
Attilio Cucchieri, Tereza Mendes
%Numerical test of the Gribov-Zwanziger scenario in Landau gauge
arXiv:1001.2584 [hep-lat].


\bibitem{Cucchieri:2008.12}
Attilio Cucchieri, Tereza Mendes,
%%Infrared behavior and infinite-volume limit of gluon and ghost propagators in Yang-Mills theories%%
 arXiv:0812.3261 [hep-lat].   

\bibitem{Cucchieri:2008.04}
Attilio Cucchieri, Tereza Mendes,
%%Constraints on the IR behavior of the ghost propagator in Yang-Mills theories%%
Phys.~Rev.~{\bf D78} (2008) 094503 and arXiv: 0804.2371 [hep-lat].


\bibitem{Ilgenfritz:2009}
I.L. Bogolubsky, E.-M. Ilgenfritz, M. MŸller-Preussker, A. Sternbeck,
%%Lattice gluodynamics computation of Landau-gauge Green's functions in the deep infrared
arXiv:0901.0736 [hep-lat].


\bibitem{Oliveira:2008}
O. Oliveira, P. J. Silva,
%%Does The Lattice Zero Momentum Gluon Propagator for Pure Gauge SU(3) Yang-Mills Theory Vanish in the Infinite Volume Limit?
arXiv:0809.0258 [hep-lat].
  


\bibitem{vonSmekal2007}
A. Sternbeck, L. von Smekal, D. B. Leinweber, A. G. Williams
%%Comparing SU(2) to SU(3) gluodynamics on large lattices
PoS {\bf LAT2007} 340, 2007 arXiv:0710.1982 [hep-lat].
  
  
\bibitem{Cucchieri:2007b}  
A. Cucchieri, T. Mendes,
%%Constraints on the IR behavior of the gluon propagator in Yang-Mills theories
Phys. Rev Lett. {\bf 100} 241601, 2008 and arXiv:0712.3517 [hep-lat].

\bibitem{Gribov:1977wm}
V. N. Gribov,
%%\textit{Quantization of non-Abelian Gauge Theories
Nucl. Phys. B {\bf 139}, 1 (1978).

\bibitem{Zwanziger:1991}
Daniel Zwanziger
%%Vanishing of zero-momentum lattice gluon propagator and color confinement
Nucl. Phys. B {\bf 364}, 127 (1991).


\bibitem{Maas:2007} 
A.~Maas,
%%Two- and three-point Green's functions in two-dimensional Landau-gauge Yang-Mills theory
Phys. Rev. {\bf D75} 116004 (2007) and arXiv:0704.0722 [hep-lat].










\end{thebibliography}
\end{document}